\documentclass[aps,prc,reprint, showpacs,superscriptaddress]{revtex4-1}
\usepackage{graphicx}
\usepackage{dcolumn}
\usepackage{bm}
\begin{document}
\title{Fission Fragment Angular Anisotropy in Neutron-Induced Fission of $^{235}$U Measured with a Time Projection Chamber}


\author{V.~Geppert-Kleinrath}\email{verena@lanl.gov}\affiliation{Los Alamos National Laboratory, Los Alamos, New Mexico 87545}
\author{F.~Tovesson}\affiliation{Los Alamos National Laboratory, Los Alamos, New Mexico 87545}
\author{J.S.~Barrett}\affiliation{Oregon State University, Corvallis, Oregon 97331}
\author{N.S.~Bowden}\affiliation{Lawrence Livermore National Laboratory, Livermore, California 94550}
\author{J.~Bundgaard}\affiliation{Colorado School of Mines, Golden, Colorado 80401}
\author{R.J.~Casperson}\affiliation{Lawrence Livermore National Laboratory, Livermore, California 94550}
\author{D.A.~Cebra}\affiliation{University of California, Davis, California 95616}
\author{T.~Classen}\affiliation{Lawrence Livermore National Laboratory, Livermore, California 94550}
\author{M.~Cunningham}\affiliation{Lawrence Livermore National Laboratory, Livermore, California 94550}
\author{D.L.~Duke}\affiliation{Los Alamos National Laboratory, Los Alamos, New Mexico 87545}
\author{J.~Gearhart}\affiliation{University of California, Davis, California 95616}
\author{U.~Greife}\affiliation{Colorado School of Mines, Golden, Colorado 80401}
\author{E.~Guardincerri}\affiliation{Los Alamos National Laboratory, Los Alamos, New Mexico 87545}
\author{C.~Hagmann}\affiliation{Lawrence Livermore National Laboratory, Livermore, California 94550}
\author{M.~Heffner}\affiliation{Lawrence Livermore National Laboratory, Livermore, California 94550}
\author{D.~Hensle}\affiliation{Colorado School of Mines, Golden, Colorado 80401}
\author{D.~Higgins}\affiliation{Los Alamos National Laboratory, Los Alamos, New Mexico 87545}\affiliation{Colorado School of Mines, Golden, Colorado 80401}
\author{L.D.~Isenhower}\affiliation{Abilene Christian University, Abilene, Texas 79699}
\author{J.~King}\affiliation{Oregon State University, Corvallis, Oregon 97331}
\author{J.L.~Klay}\affiliation{California Polytechnic State University, San Luis Obispo, California 93407}
\author{W.~Loveland}\affiliation{Oregon State University, Corvallis, Oregon 97331}
\author{J.A.~Magee}\affiliation{Lawrence Livermore National Laboratory, Livermore, California 94550}
\author{B.~Manning}\affiliation{Los Alamos National Laboratory, Los Alamos, New Mexico 87545}
\author{M.P.~Mendenhall}\affiliation{Lawrence Livermore National Laboratory, Livermore, California 94550}
\author{J.~Ruz}\affiliation{Lawrence Livermore National Laboratory, Livermore, California 94550}
\author{S.~Sangiorgio}\affiliation{Lawrence Livermore National Laboratory, Livermore, California 94550}
\author{K.T.~Schmitt}\affiliation{Los Alamos National Laboratory, Los Alamos, New Mexico 87545}
\author{B.~Seilhan}\affiliation{Lawrence Livermore National Laboratory, Livermore, California 94550}
\author{L.~Snyder}\affiliation{Lawrence Livermore National Laboratory, Livermore, California 94550}
\author{A.C.~Tate}\affiliation{Abilene Christian University, Abilene, Texas 79699}
\author{R.S.~Towell}\affiliation{Abilene Christian University, Abilene, Texas 79699}
\author{N.~Walsh}\affiliation{Lawrence Livermore National Laboratory, Livermore, California 94550}
\author{S.~Watson}\affiliation{Abilene Christian University, Abilene, Texas 79699}
\author{L.~Yao}\affiliation{Oregon State University, Corvallis, Oregon 97331}
\author{W.~Younes}\affiliation{Lawrence Livermore National Laboratory, Livermore, California 94550}

\collaboration{NIFFTE Collaboration}\homepage{http://niffte.calpoly.edu/}\noaffiliation

\author{H.~Leeb}\affiliation{Technische Universit\"{a}t Wien, Vienna, Austria}


\date{\today}

\begin{abstract}
Fission fragment angular distributions can provide an important constraint on fission theory, improving predictive fission codes, and are a prerequisite for a precise ratio cross section measurement. Available anisotropy data is sparse, especially at neutron energies above 5 MeV. For the first time, a three-dimensional tracking detector is employed to study fragment emission angles and provide a direct measurement of angular anisotropy. The Neutron Induced Fission Fragment Tracking Experiment (NIFFTE) collaboration has deployed the fission time projection chamber (fissionTPC) to measure nuclear data with unprecedented precision. The fission fragment anisotropy of $^{235}$U has been measured over a wide range of incident neutron energies from 180 keV to 200 MeV; a careful study of the systematic uncertainties complement the data.”
 
\end{abstract}


\maketitle

\section{Introduction}
Nuclear fission is a highly complex many-body quantum mechanics problem. The dynamics of the nucleus on the path to scission are difficult to calculate and after over 75~years of fission research a lot is still to be discovered in order to fully characterize the process. 
There has been a recent resurgence in efforts to further the understanding of nuclear fission, which could provide predictive power for systems inaccessible to measurements and support all nuclear applications as well as model development \cite{Chadwick2011}.
The careful study of fission fragment angular distributions presented here has a two-fold impact: For one the data will directly support the effort to measure precision fission cross sections, a much needed nuclear data quantity for applications. Secondly, the results can aid in the further development of predictive fission models.
Measurements of cross section ratios are a common practice in the field of nuclear physics; in a ratio experiment the target of interest is measured with respect to a well-known standard cross section. As angular distributions of fission fragments can vary greatly between different target isotopes and across neutron energies, knowledge of the different distributions is required to correct for detection efficiency in the cross section ratio. Recent precision measurements of cross section ratios \cite{u5u8, Salvador-Castineira2015, Paradela2015}  use fission fragment angular distribution data for the efficiency correction to improve accuracy of the obtained ratios.

Fission fragment angular distributions also allow theorists valuable insights into the fission process. The angular distributions provide information on the state of the transition nucleus at the saddle point of the fission barrier \cite{Vandenbosch1973}. The nuclear configurations and shapes on the maximum of the fission barrier before the nucleus splits into two fragments are in principle completely inaccessible to measurement, but the angular distributions hint at those configurations \cite{Loveland1967,Behkami1968}. The presented angular anisotropy measurement can therefore be used to constrain predictive fission codes like Los Alamos National Laboratory's AVXSF (average cross section fission) program \cite{Lynn1973,Bouland2013}, a statistical code to predict fission cross sections using a Hauser-Feshbach formalism. The transition states are calculated using single particle excitations, and angular distributions can provide constraints on the transition state parameters and therefore improve the cross section prediction capability; developers are currently working on the implementation in AVXSF \cite{LynnPriv}.
\subsection{Theory}\label{sec:intro}
Angular distributions have famously driven theory development in the past. It was only in the early 1950's that anisotropic emission was first observed in photofission of $^{232}$Th \cite{Winhold1952}. It is now known thanks to Bohr's 1956 theory \cite{Bohr1956} that, due to the population of only a few discrete energy levels of the transition nucleus at the saddle point, fission fragment emission at intermediate energies is anisotropic.  

Bohr first suggested that for excitation energies near the barrier the nucleus is thermodynamically cold  as most energy is stored in deformation energy. He introduced the name transition nucleus for the deformed nucleus at the top of the barrier, which exhibits a discrete spectrum of excitations similar to the one found at the ground state deformation. Bohr's concept of near-barrier fission through discrete low-lying transition states provides the framework in which the angular distributions depend on the quantum numbers of the discrete transition states involved.
The transition state model immediately provided a qualitative understanding of anisotropic emission of fission fragments for photofission and other low energy fission reactions \cite{Griffin1965}.
Halpern and Strutinsky introduced a statistical treatment that generalizes the energy dependence of fragment angular distributions to moderate and high excitation energies above the fission barrier. This standard saddle-point statistical model \cite{Halpern1958} describes the distribution of states at the saddle point with a Gaussian distribution related to the thermodynamic temperature of the excited nucleus.

Within the framework of the statistical model one can phenomenologically explain changes in the magnitude of anisotropic emission with respect to neutron energy. One of the characteristic properties of the transition nucleus in the statistical saddle point model is its thermodynamical temperature, the energy available to populate different level states. In a simplified picture the fissioning nucleus will exhibit anisotropic emission when it is thermodynamically cold and only very few energy levels are populated. If the available energy increases, more states are populated and the emission becomes more isotropic. The anisotropy would exhibit a plateau or even decrease with higher energy until in the process of multi-chance fission a neutron is emitted which carries away excitation energy, when the now thermodynamically colder nucleus populates fewer states and exhibits a sharp rise in anisotropy. Multi-chance fission denotes the process of emission of one or more neutrons before the compound nucleus scissions and constitutes a major contribution to the neutron-induced fission cross section of $^{235}$U. The concept explains structure in anisotropy at the multi-chance fission thresholds, much like the step-like structure of the fission cross section. 
\subsection{Angular Anisotropy}
The anisotropy is a first-order descriptor of angular distributions, where the anisotropy A is defined as the fragment emission probability in neutron beam direction $W[0^{\circ}]$ over emission probability perpendicular to beam $W[90^{\circ}]$
\begin{equation}
\label{eq:ani}
A = \frac{W[0^{\circ}]}{W[90^{\circ}]}.
\end{equation}
Although the anisotropy parameter itself does not provide complete information on the shape of the angular distribution, it is a good indicator especially in the case of neutron-induced fission of $^{235}$U where anisotropic emission happens mainly in the forward-backward direction. The anisotropy parameter also allows easy comparison of data sets and isotopes over a range of incident energies.
To extract anisotropy or distribution shapes, the angular distributions $W[\theta]$ are commonly fitted with a sum of Legendre polynomials of the form 
\begin{equation}
\label{eq:legendre}
W[\theta] =  \sum_{n} c_n P_{2n}(\cos[\theta]),
\end{equation}
where $P_{2n}$ only includes even Legendre polynomials to conserve forward/backward symmetry \cite{Vandenbosch1973}.
Angular distributions are displayed in $\cos(\theta)$ space, where an isotropic distribution is flat.
\subsection{Previous Anisotropy Data}
\begin{figure}[htbp]
\includegraphics[width=1.\linewidth]{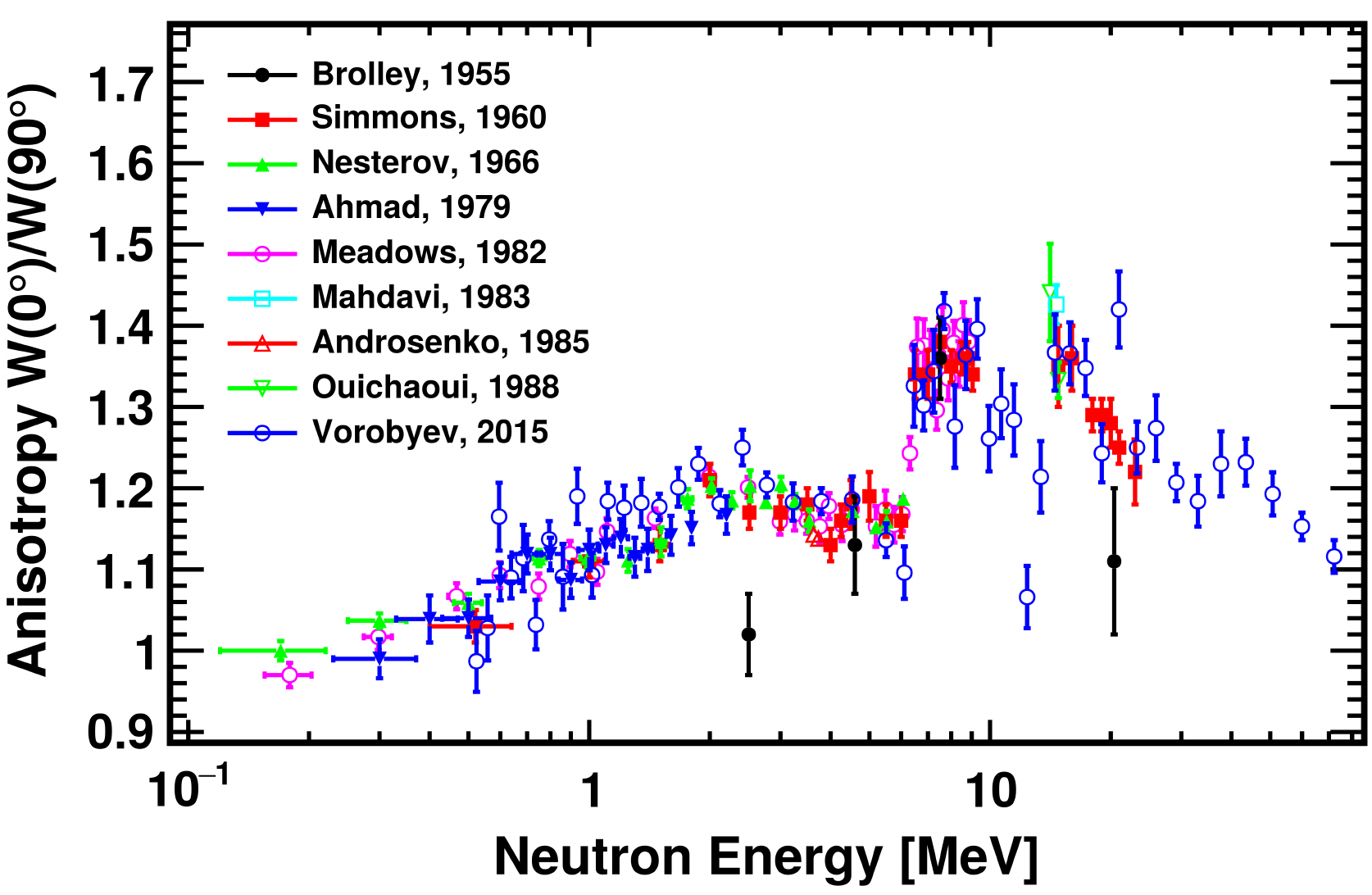}
\caption[]{(Color online) Fission fragment anisotropy data for neutron-induced fission of $^{235}$U from the EXFOR database \cite{Otuka2014}.}
\label{fig:motivation}
\end{figure}
Although the effect has been observed for some time, data on fission fragment angular distributions available in the literature is limited, as is evident in the EXFOR database \cite{Otuka2014}.  Since angular information on fission fragments is difficult to obtain in detectors that have been used to study fission in the past, most commonly ionization chambers, only a handful of measurements have been carried out. Figure \ref{fig:motivation} shows an overview of anisotropy measurements of $^{235}$U(n,f) found in the EXFOR library. 
Existing measurements agree well at neutron energies up to 10~MeV, the two most comprehensive sets in this energy region are from Simmons and Meadows \cite{Simmons1960,Meadows1983}. Vorobyev \cite{Vorobyev2015} recently added a measurement for neutron energies up to 200~MeV, but shows large scattering in the data at lower energies. Vorobyev's data were obtained at the GNEIS facility \cite{Abrosimov1985}, the only other measurement besides the one presented here carried out at a continuous neutron source.
The result described in this paper constitutes the first measurement of fission fragment angular anisotropy with a three-dimensional tracking detector, providing data over a wide range of incident neutron energies up to 200~MeV.

\section{Experimental Method}
\begin{figure}[htbp]
\fbox{\includegraphics[width=0.8\linewidth]{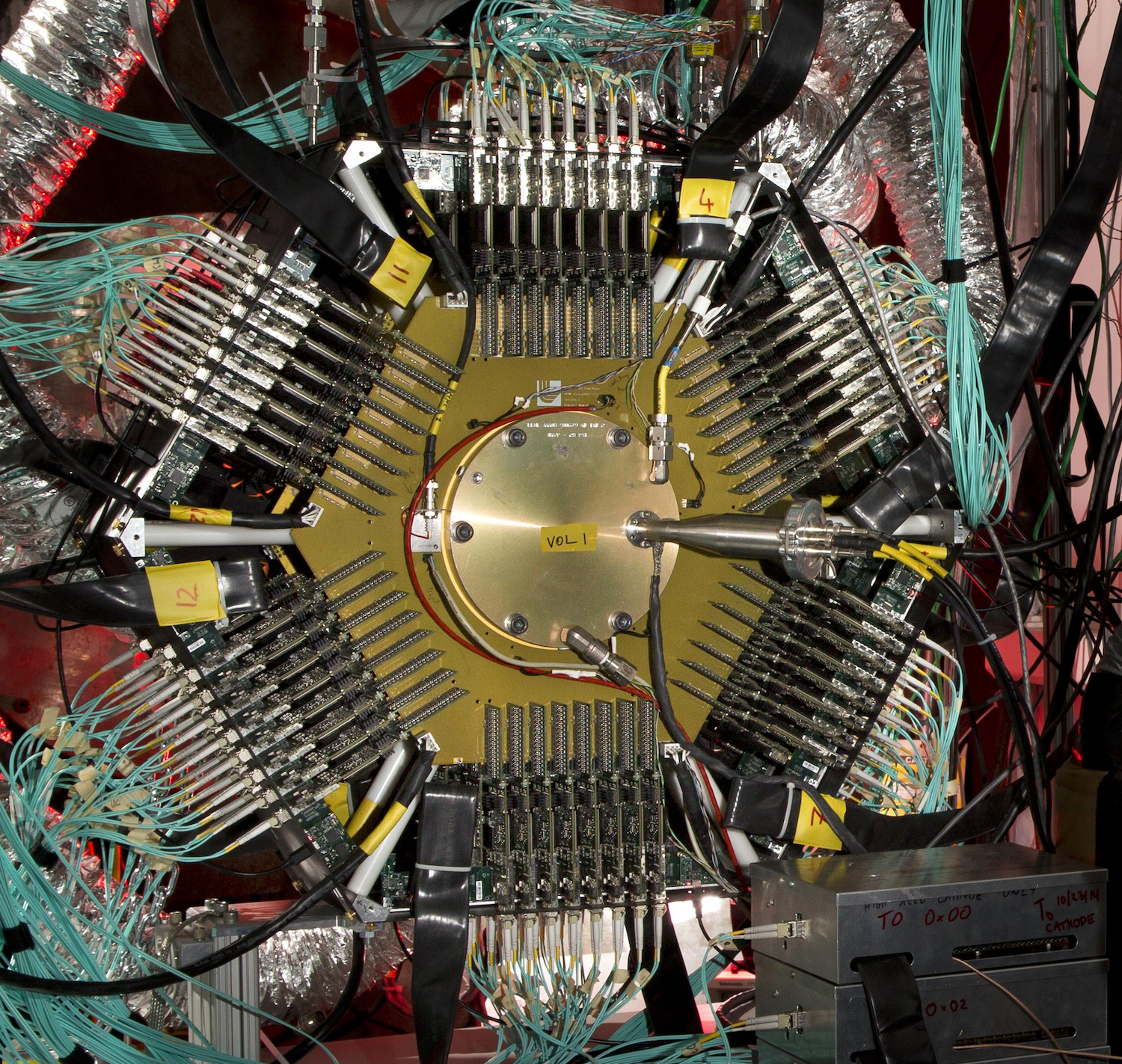}} 
\caption{(Color online) A photograph of the fully assembled fissionTPC on the 90L flightpath at LANSCE during the 2014/2015 run cycle.}
\label{fig:TPC90l}
\end{figure}

The fission time projection chamber (fissionTPC) is a time projection chamber specifically developed to study the fission process \cite{Heffner2014new}. The Neutron Induced Fission Fragment Tracking Experiment (NIFFTE) collaboration recently added this new capability to the toolbox of nuclear fission research. The fissionTPC provides three-dimensional tracking of all ionizing particles produced in the fission process. It consists of a cyclindrical pressure vessel of 7.5~cm radius and 11~cm length filled with a 95\% Argon/ 5\% Isobutane gas mixture at 550~Torr, two pixelated MICROMEGAS \cite{Giomataris1996} anode planes with 2976~hexagonal channels each, a cathode holding the actinide target in the center, a field cage providing a stable electric field of 520~V/cm, and ethernet-based data acquisition electronics. Figure \ref{fig:TPC90l} shows a photograph of the fissionTPC fully assembled on the 90L flightpath at the Los Alamos Neutron Science Center (LANSCE) \cite{Lisowski2006}.

The LANSCE facility houses an 800~MeV linear proton accelerator, which produces neutrons at two different spallation targets \cite{Lisowski1990}. Target-4 is a high-energy white neutron source that produces neutrons at energies ranging from 0.1~MeV to greater than 600~MeV. Proton pulses of less than 250~ps width impinge on the tungsten spallation target at 100~Hz with an average current of 1.5~$\mu$A, producing neutron macropulses of 625~$\mu$s width. Each macropulse contains a succession of 1.8~$\mu$s long micropulses, where the spacing can be extended to 3.6~$\mu$s to investigate neutron background. Because of the pulsed nature of the proton and subsequently neutron beam, neutron energy measurement via time of flight is possible on an event by event basis.

A 0.5~mm thick backing aluminium target with 20~mm diameter actinide deposits of 50~$\mu$g/cm$^2$ $^{235}$U on one side and $^{239}$Pu on the other side was used in this work. The isotopic purity of the $^{235}$U deposit was confirmed through alpha spectrometry to be over 99.5\%. 
Destructive mass spectrometry confirmed that trace contaminants of $^{233}$U, $^{234}$U,  $^{236}$U and $^{238}$U contributed less than 0.5$\%$ to the target atoms. As a result, less than 0.5$\%$ of the fission fragments arise from these contaminants which have fission cross-sections of the same order of magnitude as  $^{235}$U. The uncertainty on the anisotropy resulting from these contaminants was therefore neglected.

\section{Data Analysis}
The first step in the analysis is the reconstruction of the raw signals from the fissionTPC, transforming them into three-dimensional particle tracks with energy loss information. The two-dimensional (x,y) coordinates of the ionizing particle track in the detector volume stem from the position of each pad receiving ionization electrons on the segmented anode plane. The z-coordinate is based upon the arrival time of the signal and is correlated to the drift speed of the ionization electrons within the fissionTPC. The absolute z-position of tracks in the fissionTPC can be determined by requiring tracks to originate from the cathode plane. 
The height of the signal is correlated to the number of ionization electrons deposited on the pad plane and therefore to the energy of the ionizing particle that generated the track in the active volume of the fissionTPC. The reconstruction software relies on the individual signal shapes, the signals' channel locations on the pad plane, their time-stamp, and the shape and size of their rise. The absence of a magnetic field greatly simplifies the track extraction because it implies only straight tracks.
For the angular anisotropy analysis, tracks are found using a conventional track finder based on charge thresholds and a fitter that minimizes the distance of charge to the track. 
\subsection{Fission Fragment Identification}
\begin{figure}[htbp]
\includegraphics[width=1\linewidth]{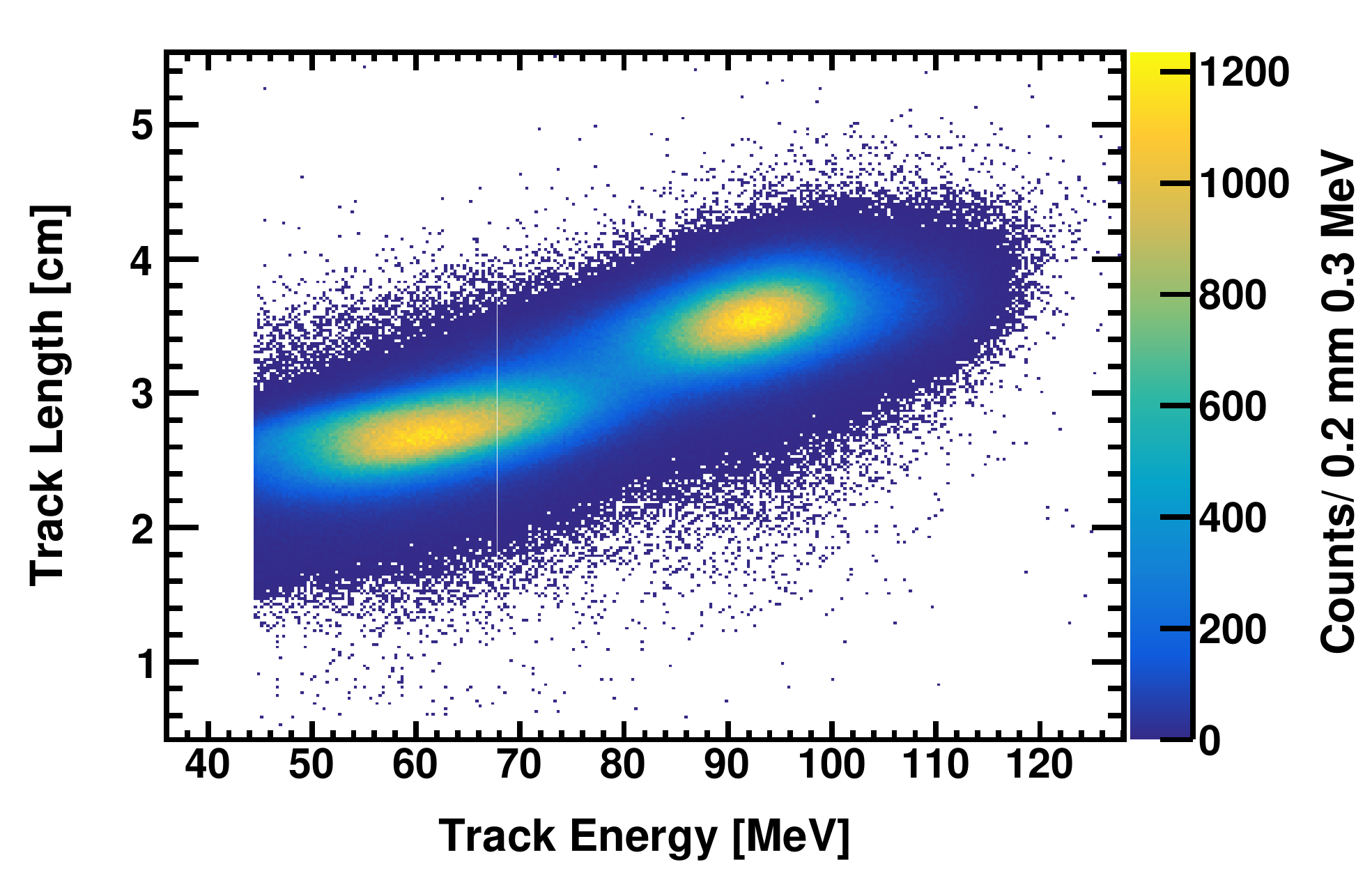}
\caption[]{(Color online) Track length versus energy for all fission fragments selected for the anisotropy analysis. The double humped distribution is characteristic of fission.}
\label{fig:selectedData}
\end{figure}

Utilizing the three-dimensional reconstruction capability of the fissionTPC one can distinguish fission fragments from other particles using charge and specific ionization information. Analysis also provides the origin of a particle track, determines polar angle, track length, and energy deposited along the track. In this analysis a total energy selection cut is placed on the detected particle tracks at 44.4~MeV to distinguish fission from recoil ions and other particles. The implications of placing the cut at that specific energy are discussed in section \ref{sec:unc}. Figure \ref{fig:selectedData} shows the selected data associated with fission fragments in a length versus energy plot.
\subsection{Neutron Time of Flight Calibration}
\begin{figure}[htbp]
\includegraphics[width=1\linewidth]{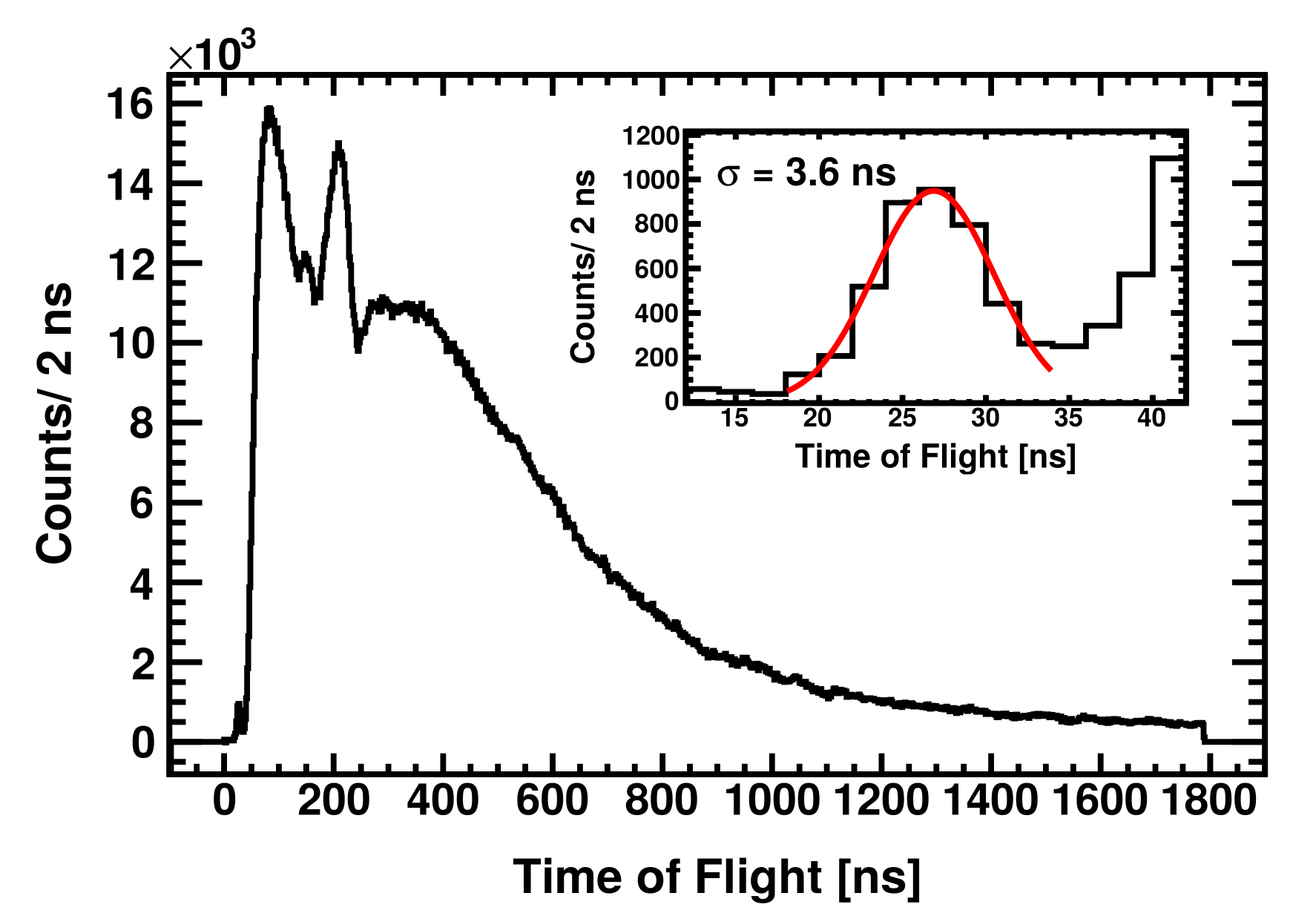}
\caption[]{(Color online) The neutron time of flight spectrum for the selected fission fragments passing the energy cut selection. The inset shows the fitted photofission peak with a standard deviation $\sigma$ = 3.6~ns.}
\label{fig:nTOF_photofiss}
\end{figure}

Each proton micropulse produces a range of neutron energies, where the specific energy causing the fission event can be determined based upon the neutron flight time between the spallation target and the fissionTPC in the experimental area. The two signals to determine the time of flight are the accelerator signal which relates to the neutron pulse production time and the fissionTPC cathode signal for each fission event.
An important feature of the TOF spectrum is the photofission peak at small times of flight, shown in the inset in Figure \ref{fig:nTOF_photofiss}. 
The peak indicates the TOF of speed of light particles (gamma-rays produced in the spallation process) from which the TOF distance can be calculated. 
Placing a carbon filter in  the beam, a resonance at 2078$\pm$0.3~keV in the carbon total cross section \cite{Carbon} produces a notch in the TOF spectrum. From the position of the notch relative to the photofission peak one can calculate the flight path length, which was 8.028$\pm$0.015~m.

Since the proton beam is a narrow packet of less than 250~ps, the gamma time of flight is well defined and the photofission peak width ($\sigma$ = 3.6~ns) is dominated by the detector timing resolution. 
The uncertainty in timing from the photofission peak resolution is carried through when calculating neutron energies, which results in an energy uncertainty for each energy. The bin width is chosen to be larger than twice the uncertainty in energy. The binning for the anisotropy result is linear to 1~MeV and then logarithmic with 14~bins/decade up to 20~MeV and 7~bins/decade at higher energies.
\subsection{Efficiency Normalization}
\begin{figure}[htbp]
\includegraphics[width=0.9\linewidth]{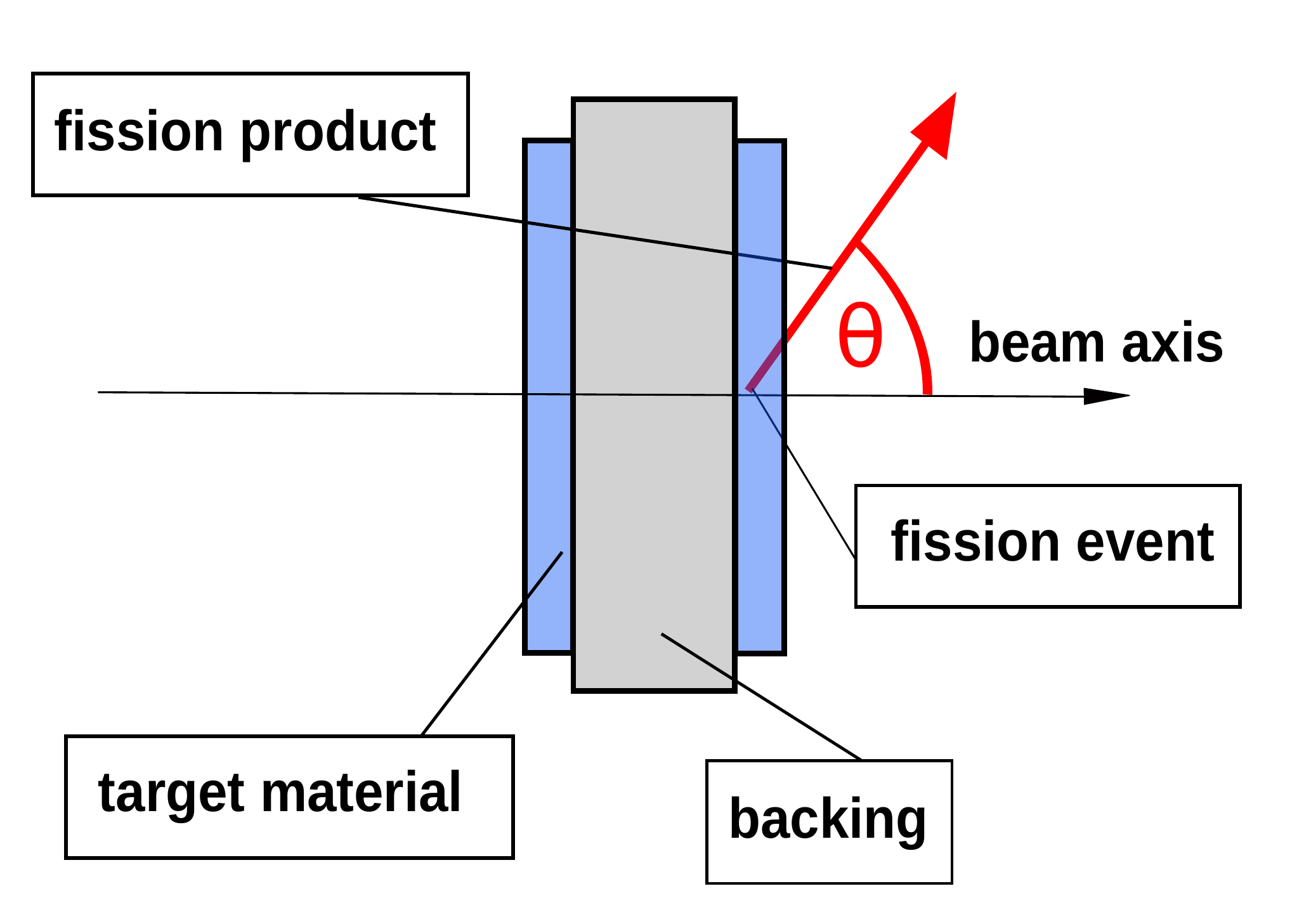}
\caption[]{(Color online) The thick backing target with actinide deposits on each side and a fission fragment leaving the target material. The geometry determines the efficiency for detecting fission fragments in the fissionTPC.}
\label{fig:geometry}
\end{figure}

\begin{figure}[htbp]
\includegraphics[width=1\linewidth]{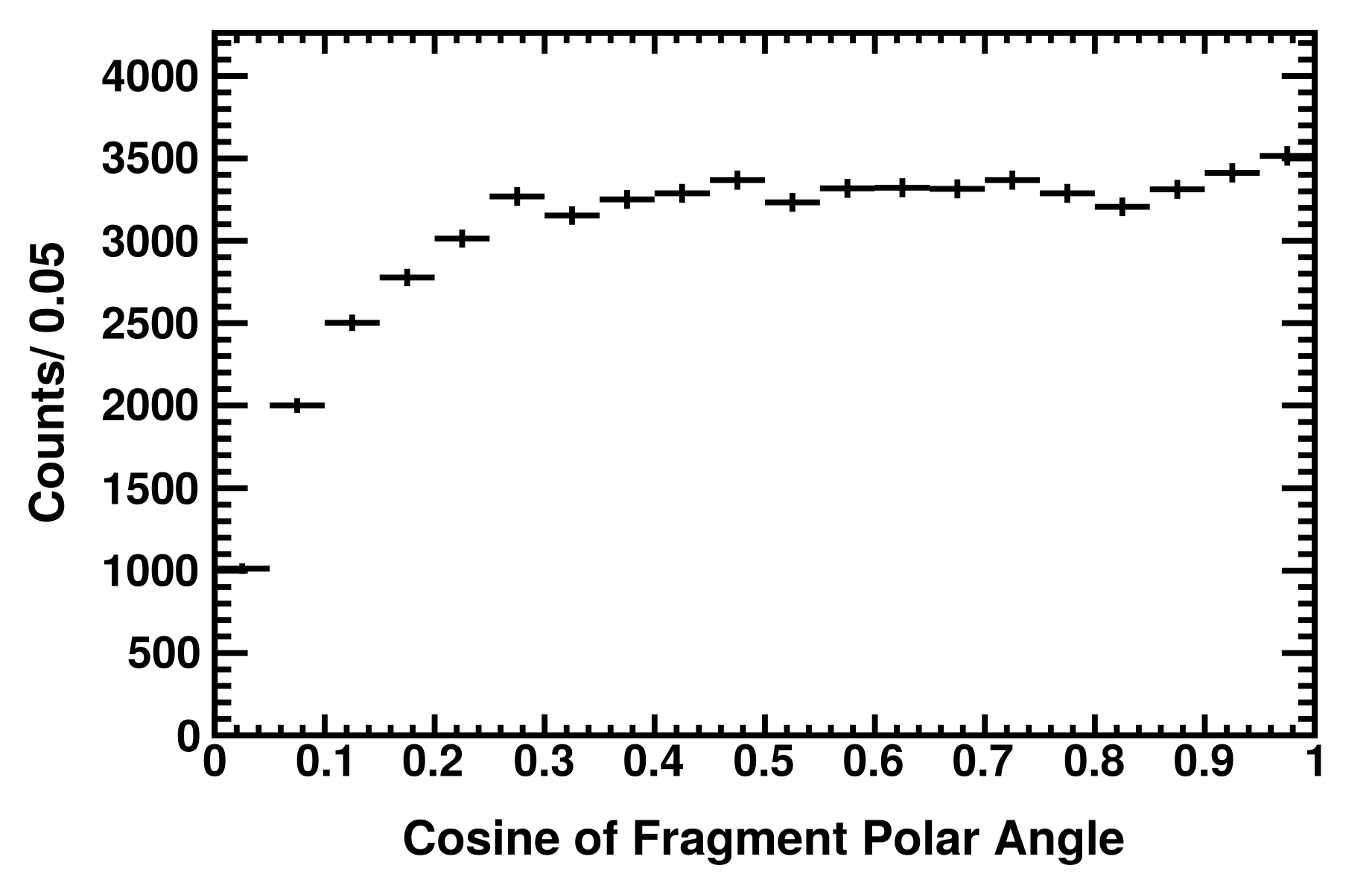}
\caption[]{The angular distribution of fission fragments in the fissionTPC for incident neutron energies from 130~keV to 160~keV, where emission is assumed to be isotropic. A steep drop-off in efficiency close to $\cos(\theta) = 0$ is observed.}
\label{fig:norm_dist}
\end{figure}

The angular distributions are normalized for detection efficiency. The angular dependent efficiency is a consequence of the target geometry in the fissionTPC, target homogeneity and beam characteristics, as described in more detail in \cite{u5u8}. Figure \ref{fig:geometry} shows a schematic depiction of the thick backing target with the actinide deposits on each side.  If fission fragments are emitted at an angle parallel to the target they will fail to escape the material, meaning that the efficiency for detecting fragments drops off steeply at $\cos(\theta) = 0$ (parallel to the target surface). Figure \ref{fig:norm_dist} shows the measured angular distribution in the lowest available energy bin from 130~keV to 160~keV. If we assume that the physical angular distribution of fission fragments is isotropic at these energies as discussed in section \ref{sec:unc}, the distribution shown in Figure \ref{fig:norm_dist} reflects the measured efficiency of the detector.
With the additional assumption that the obtained efficiency is neutron energy independent the measured angular distributions in all energy bins are divided bin-by-bin through the efficiency curve from Figure \ref{fig:norm_dist} and normalized so that the Legendre polynomial fit described in section \ref{sec:extraction} $W[\cos(\theta)=0] = 1$ to allow visual comparisons between different distributions. This procedure yields the efficiency-corrected physical angular distributions. The counting statistics error bars in the measured angular distributions are carried through the arithmetic operations.
\subsection{Linear Momentum Transfer}
Although the mass of the incident neutron is considerably smaller than the mass of the actinide, it transfers some momentum to the target nucleus and subsequently to the fission fragments. Forward fragments will receive a boost in energy and therefore an enhancement at forward angles will be observed, while the opposite is true for backwards fragments. The change in kinetic energy of fission fragments and their angles can be calculated from conservation of momentum and energy in the center-of-mass system of incident neutron and actinide target atom, see a detailed derivation in reference \cite{kleinrath}. The kinematic quantities are computed in a classical, non-relativistic framework. The resulting angular distributions used for the derivation of anisotropy are then given in the center-of-mass frame.
\subsection{Extracting Anisotropy}
\label{sec:extraction}
\begin{figure}[htbp]
\includegraphics[width=1\linewidth]{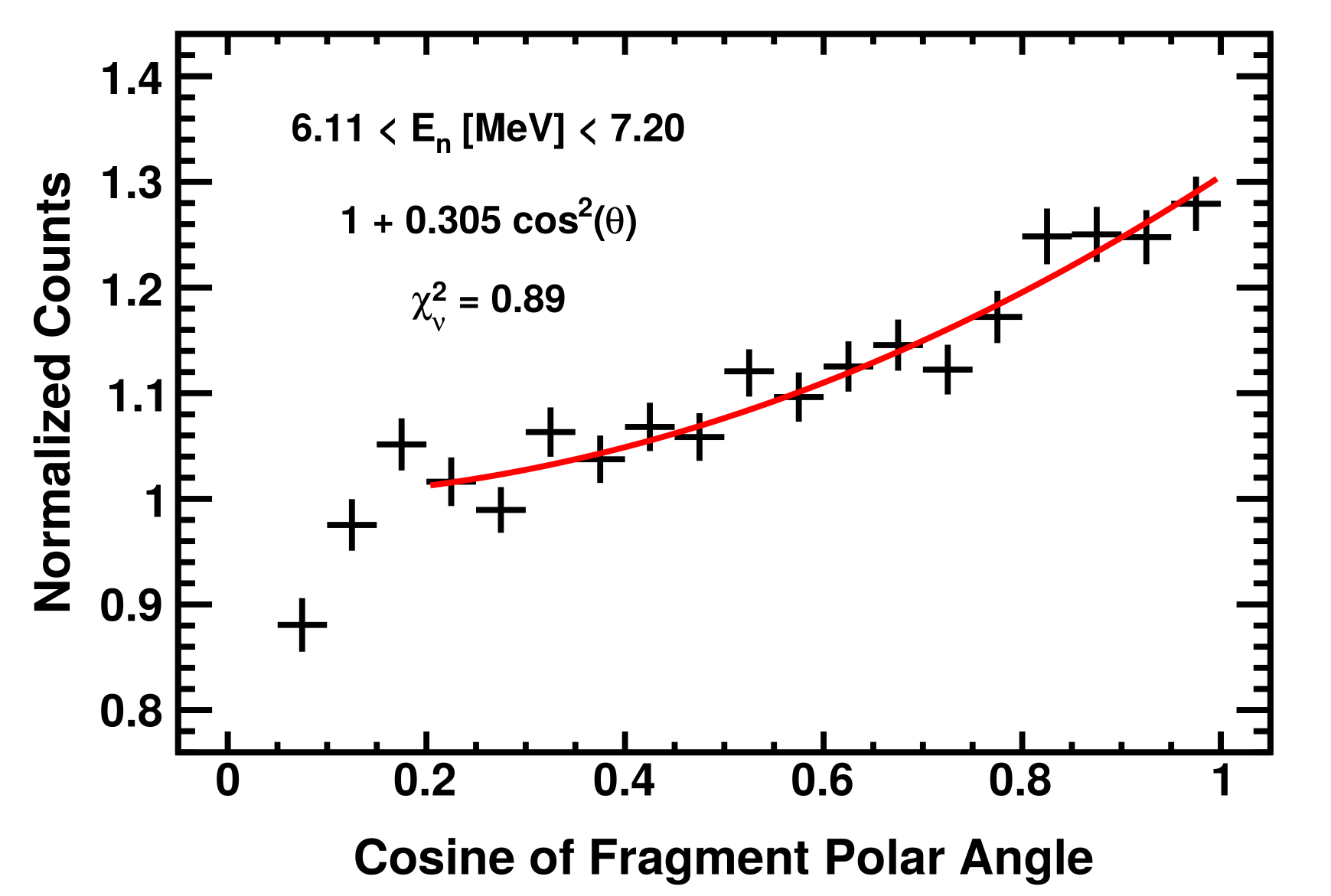} 
\caption[]{(Color online) The efficiency-corrected angular distribution for fission events initiated by 6.11-7.20~MeV neutrons, fit with second-order Legendre polynomials in $\cos(\theta)$ space.}
\label{fig:angularfits} 
\end{figure}

To extract the anisotropy, the angular distributions are fit with the Legendre polynomials in equation \ref{eq:legendre} using a least squares fitting routine. Figure \ref{fig:angularfits} shows a typical normalized angular distribution including the polynomial fit and associated goodness of fit parameter $\chi^2_{\nu}$ (reduced $\chi^2$) in the 6.11 to 7.20~MeV neutron energy bin. All angular distributions with associated fit functions are available in the supplemental material \cite{supplement}. The angular distributions are fit from $\cos(\theta) = 0.2$ to $\cos(\theta) = 1$ to omit the angles parallel to the target surface, which contain the fewest tracks and are most affected by the efficiency correction. The anisotropy is equal to the sum of the coefficients in the polynomial fit equation or the value of the fit at $\cos(\theta) = 1$, since it is normalized to 1 at $\cos(\theta) = 0$, see equations \ref{eq:ani} and \ref{eq:legendre}. The anisotropy result is derived using second order Legendre polynomials only, since including higher order terms improved the goodness of fit parameter $\chi^2_{\nu}$ only marginally while increasing the statistical uncertainty of the fit. 
\subsection{Wrap-Around Correction}
An important correction to the anisotropy result is warranted due to the structure inherent to the pulsed neutron beam at LANSCE. The spacing between micropulses is 1800~ns, the length of a single TOF spectrum. Slower neutrons with a time of flight greater than 1800~ns will still be on their way to the fissionTPC when a new pulse starts. This effect is called wrap-around or frame overlap. 
A number of overlapping tails from previous pulses lie underneath every TOF spectrum and need to be subtracted out of the final TOF. Background neutrons due to wrap-around are slower than 1800~ns, which translates to all wrap-around neutrons having an energy less than 104~keV in the fissionTPC experiment. Since we are assuming that neutrons below the efficiency normalization energy of 160~keV lead to isotropic angular distributions, the wrap-around neutrons skew the angular distribution towards isotropy and reduce the anisotropy.

The number of wrap-around neutrons is estimated by fitting the tail ends of all macropulses at neutron flight times beyond 1800~ns. The fit function used is a combination of decaying exponentials to account for several macropules,  a technique previously used for cross section measurements at LANSCE \cite{tovesson}.  The function includes a scaling parameter, which is only invoked if the background below the photofission peak is not sufficiently accounted for. The correction therefore also accounts for constant neutron background, like neutrons that scatter off equipment in the experiment room (room return).
 \begin{figure}[htbp]
\centering
\includegraphics[width=1\linewidth]{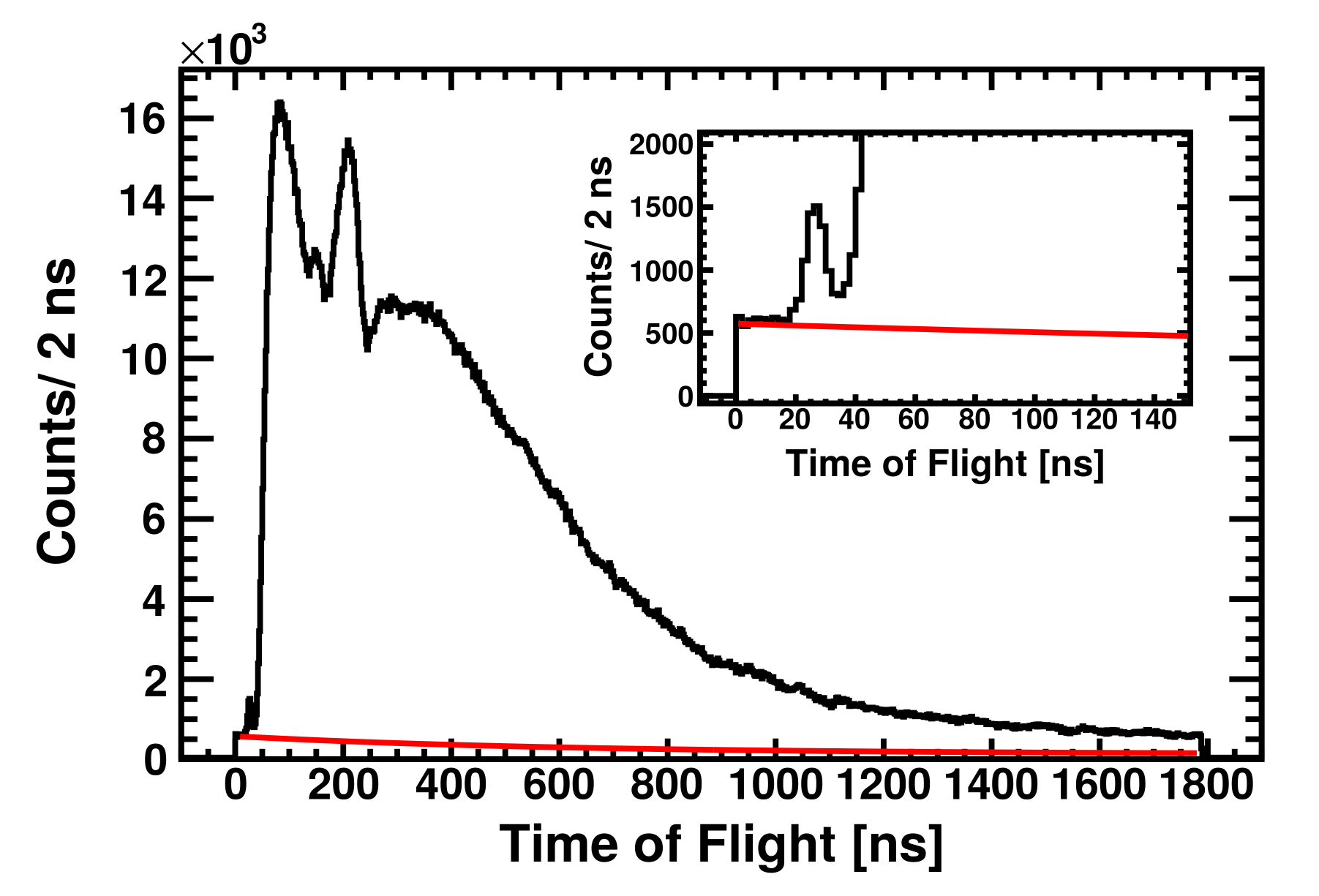}
\caption[]{(Color online) The fission fragment time of flight spectrum with the wrap-around contribution shown below. The inset shows that the wrap-around contribution is scaled to match the counts below the photofission peak.}
\label{fig:wrapfit}
\end{figure}

The obtained function is then used to determine the number of wrap-around neutrons versus the number of physical counts in each TOF bin. 
The percentage of wrap-around neutrons in each neutron energy bin is used to correct the anisotropy result, but not the individual angular distributions. The measured anisotropy value $A_{\text{meas}}$ in any particular neutron energy bin is the sum of the real anisotropy $A_{\text{real}}$ plus a flat contribution $A_{\text{wrap}}$ = 1 from wrap-around neutrons:
\begin{equation}
 A_{\text{meas}} = (1-p_{\text{wrap}}) A_{\text{real}} + p_{\text{wrap}} A_{\text{wrap}},
\end{equation}
where $p_{\text{wrap}}$ is the percentage of wrap-around neutrons in that energy bin. The corrected anisotropy value is then $A_{\text{real}}$.

\section{Uncertainty Quantification} \label{sec:unc}
The careful treatment of uncertainties related to the extraction of anisotropy values and the corrections to the data was carried out through variational analysis. Since there is no analytical relationship between the anisotropy values and manipulation of the measured data like selection cuts or fitting procedures, input parameters were varied to estimate uncertainties.

\begin{figure}[htbp]
\includegraphics[width=1\linewidth]{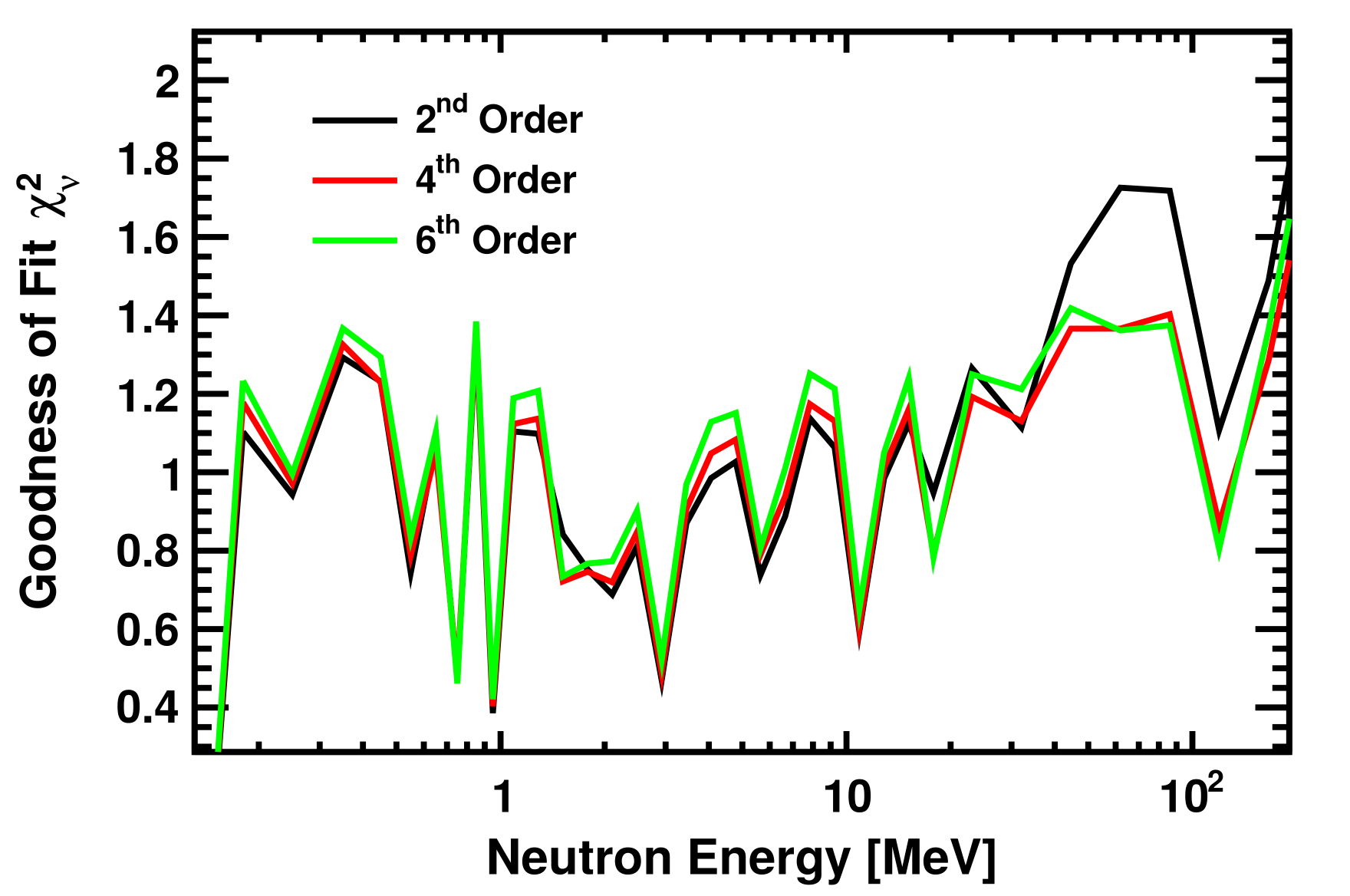}
\caption[]{(Color online) The goodness of fit parameter $\chi^2_{\nu}$ does not improve significantly with including higher order terms in the Legendre polynomial used to fit the angular distributions.}
\label{fig:chisquare}
\end{figure}

When fitting the angular distributions, including higher than quadratic terms only gives marginal improvement in the goodness of fit $\chi^2_{\nu}$ in a few neutron energy bins, see Figure \ref{fig:chisquare}, while increasing the statistical uncertainty, suggesting the quadratic fit is consistent with information available in the data. The previous anisotropy data in Figure \ref{fig:motivation} had been fit with fixed orders of 2 or 4, or the highest order of the fit had been varied with goodness of fit up to order 6 in each respective neutron energy bin. These methods have all been used to compute the anisotropy for the variational analysis and the deviation from quadratic fits is defined to be an uncertainty in the final result. The contribution is largely asymmetric, as including higher fit orders seems to systematically lower the anisotropy result, and on the order of 1 to 6~\%.

The second largest source of uncertainty is the fission fragment identification. The method of choosing fragments relies on a simple energy cut at 44.4~MeV to distinguish alpha particles from fission fragments. To deduce the uncertainty of the selection cut on the result, the energy threshold is varied in steps to a lowest value of 37~MeV where one is clearly including recoil ions and up to 52~MeV where large numbers of the lower energy fission fragments are cut out. The resulting changes in the anisotropy are largely asymmetric and the variation is on the order of 0.5 to 3~\%.

\begin{figure}[htbp]
\includegraphics[width=1\linewidth]{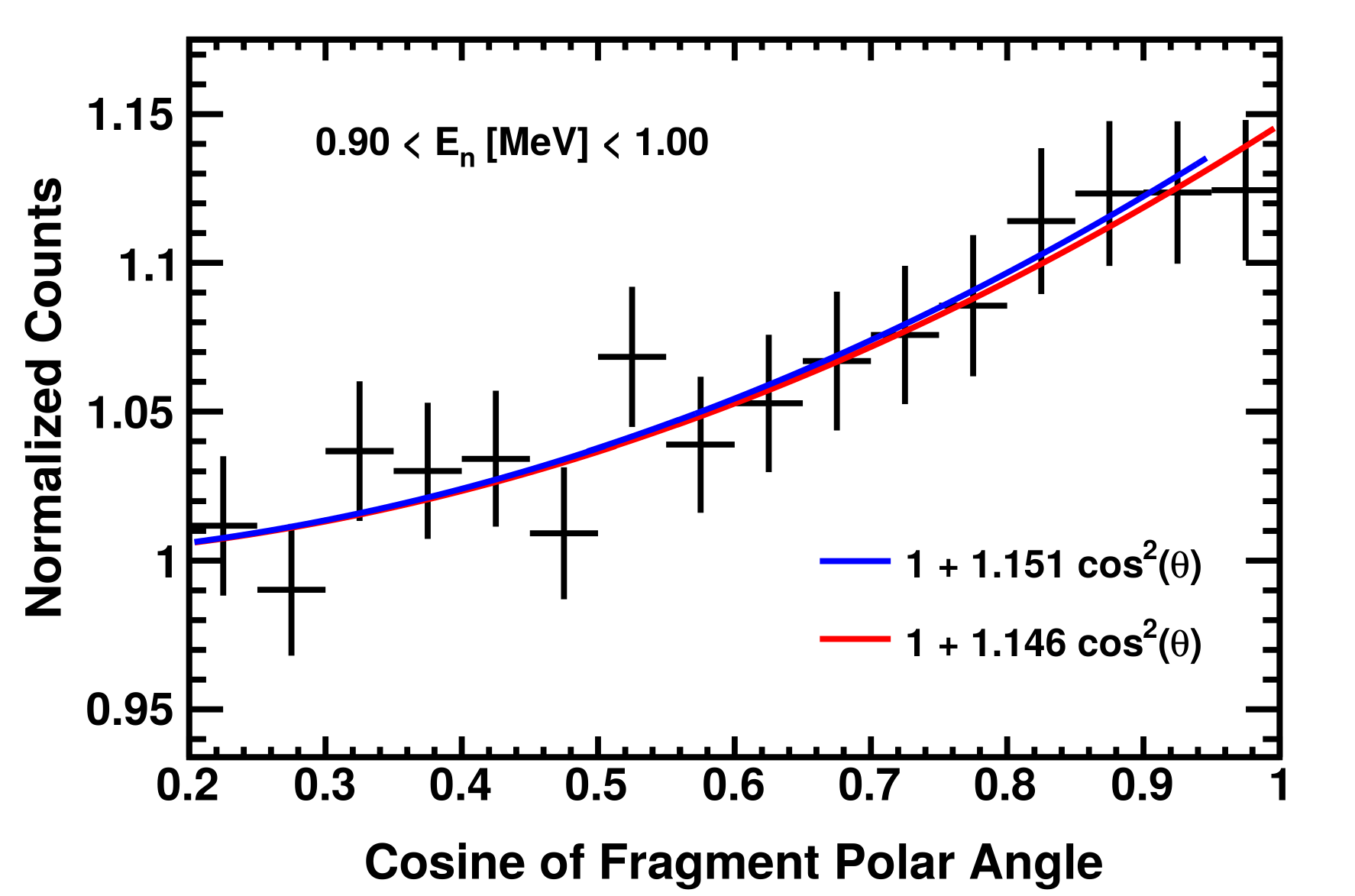}
\caption[]{(Color online) Quadratic fits for two different $\cos(\theta)$ ranges in the angular distribution from 0.9 to 1~MeV show that the anisotropy value changes for different fit ranges.}
\label{fig:range}
\end{figure}

The angular distributions are being fit from $\cos(\theta) =$ 0.2 to 1 to obtain the anisotropy result. Variational analysis using different fit ranges shows that a shorter fit range systematically reduces the anisotropy, as seen in Figure \ref{fig:range}, which shows the quadratic fit of one angular distribution for two different ranges from  $\cos(\theta) = 0.2$ to $\cos(\theta) = 0.95$ and $\cos(\theta) = 1$. Upper range limits as low as $\cos(\theta) = 0.95$ or $\theta = 18.2^{\circ}$ are considered for the uncertainty. The limited fit ranges change the result by less than 0.5 to 3~\%.

A potential bias in the three-dimensional reconstruction of particle tracks from raw data in the fissionTPC is the uncertainty in the drift speed, see \cite{gasstudy}. Changing the nominal drift speed value in the reconstruction changes the z-positions of reconstructed signals and the track length and polar angle of reconstructed fission fragment tracks. These changes are uniform over the detector volume and the angular distributions are  efficiency-normalized during the analysis, therefore the change in drift speed should not affect the final angular distributions. A variation in the result due to a change in drift speed indicates systematic uncertainties that might result from incorrect track reconstruction. Varying the drift speed from 3.12~cm/$\mu$s, as obtained with the Magboltz simulation package \cite{Biagi1999}, to 2.94~cm/$\mu$s, where the distribution of spontaneous alpha decay particles is most isotropic and therefore physical, results in a 0.5 to 1~\% uncertainty. The uncertainty in the tracking is a conservative estimate, as suggested by a detailed consideration of tracking performance for cross section measurements \cite{u5u8}.

The uncertainty introduced with the momentum transfer is a result of the assumption that has to be made on the fission fragment mass based on fragment energy to calculate the kinematics.  The mass estimate does not include any neutron emission and the momentum transfer calculation relies on measuring both fragment energies, while the current thick backing target configuration only allows the measurement of one of the fragments. The individual anode channels in the fissionTPC are not gain-calibrated yet, leading to an energy resolution of up to 15~\% of the fragment energy. The combined effect on the mass determination is conservatively estimated to approximately 15~AMU. From the large uncertainty in mass an uncertainty in the momentum transfer kinematic calculations follows, which leads to a small (less than 0.5~\% for most neutron energies) and symmetric uncertainty in the anisotropy result. 

\begin{figure}[htbp]
\includegraphics[width=1\linewidth]{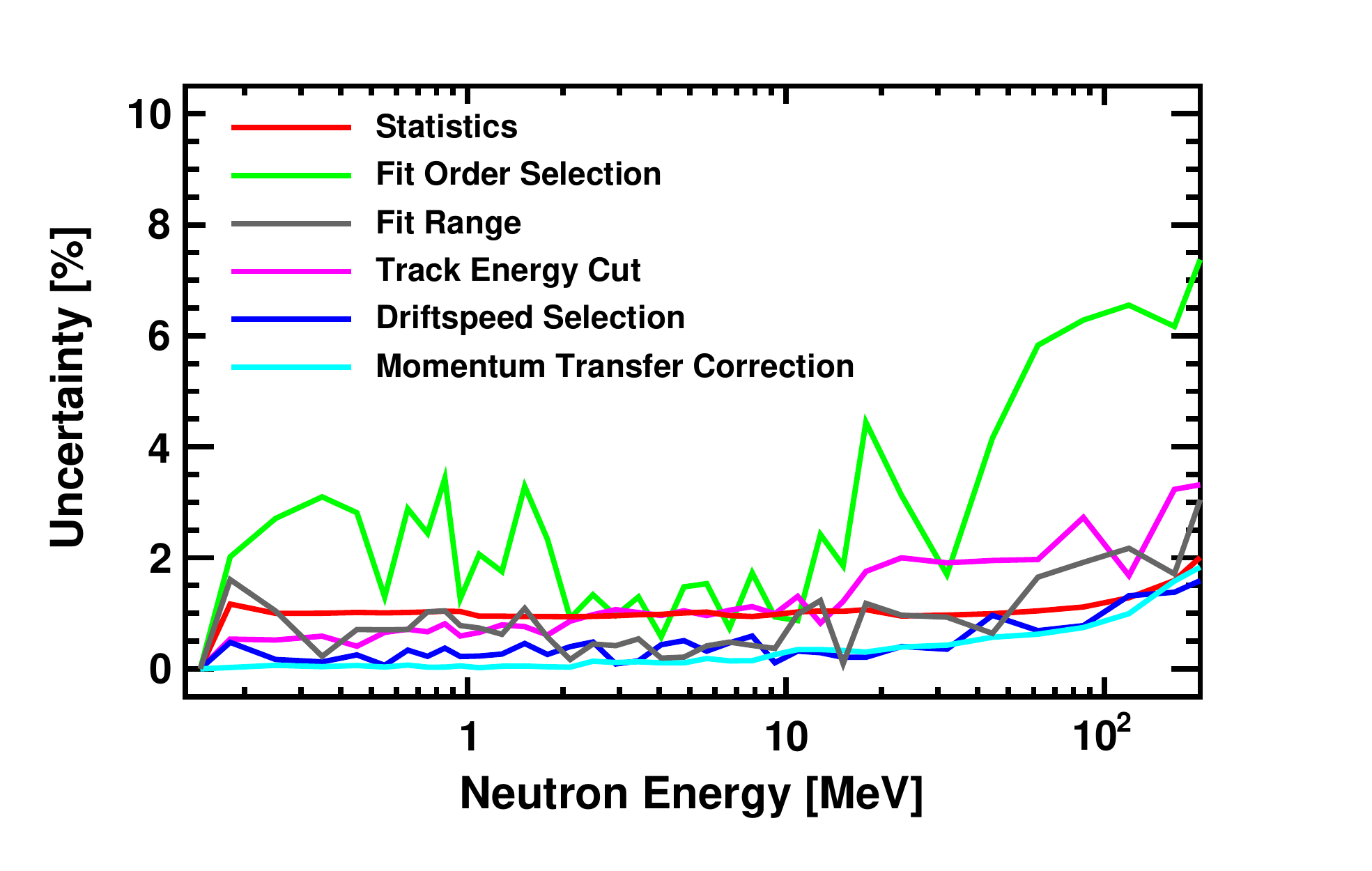}
\caption[]{(Color online) The individual uncertainty terms considered in the uncertainty study of the anisotropy result, given in percent of the anisotropy result.}
\label{fig:uncertainties}
\end{figure}

The wrap-around correction introduces a small systematic uncertainty as well. MCNP simulations of the neutron beam and the experimental facility show that the fit equation can differ from the distribution of background by up to 10~\%. Subsequently, the wrap-around term is scaled by $\pm 10\%$ to determine the result of the uncertainty on the anisotropy. The error is small (less than 0.1~\%) and symmetric around the original result and while included in the final systematic uncertainty, it is not shown in Figure \ref{fig:uncertainties}.

The statistical uncertainty, which is the uncertainty in the quadratic polynomial fit parameter,  is relatively constant at 1~$\%$ across all neutron energies. The choice of highest order of Legendre polynomial and the particle identification using a track energy cut are the largest sources of uncertainty ranging from 1 to 6~\% of the result, while the other systematics fall in the 1~\% region, as seen in Figure \ref{fig:uncertainties}. The systematic uncertainties in percent in tabular form are available in the supplemental material \cite{supplement}.

One source of variation not included in the uncertainty study is the isotropy assumption for low energy neutrons used in the efficiency normalization. While we assume that the anisotropy is 1 in the lowest energy bin from 130~keV to 160~keV, that assumption is not necessarily exact. A survey of available experimental data at low neutron energies \cite{Meadows1983,Straede1987,Musgrove,nesterov1966} shows that while most authors measured the anisotropy to be 1 or close to 1 at low energies, it could be as low as 0.95 in the lowest bin, which would systematically shift the current anisotropy result to larger values by 5~$\%$. This potential relative shift is not included in the systematic uncertainty.
\section{Results and Discussion}
\begin{figure*}[htbp]
\includegraphics[width=1\linewidth]{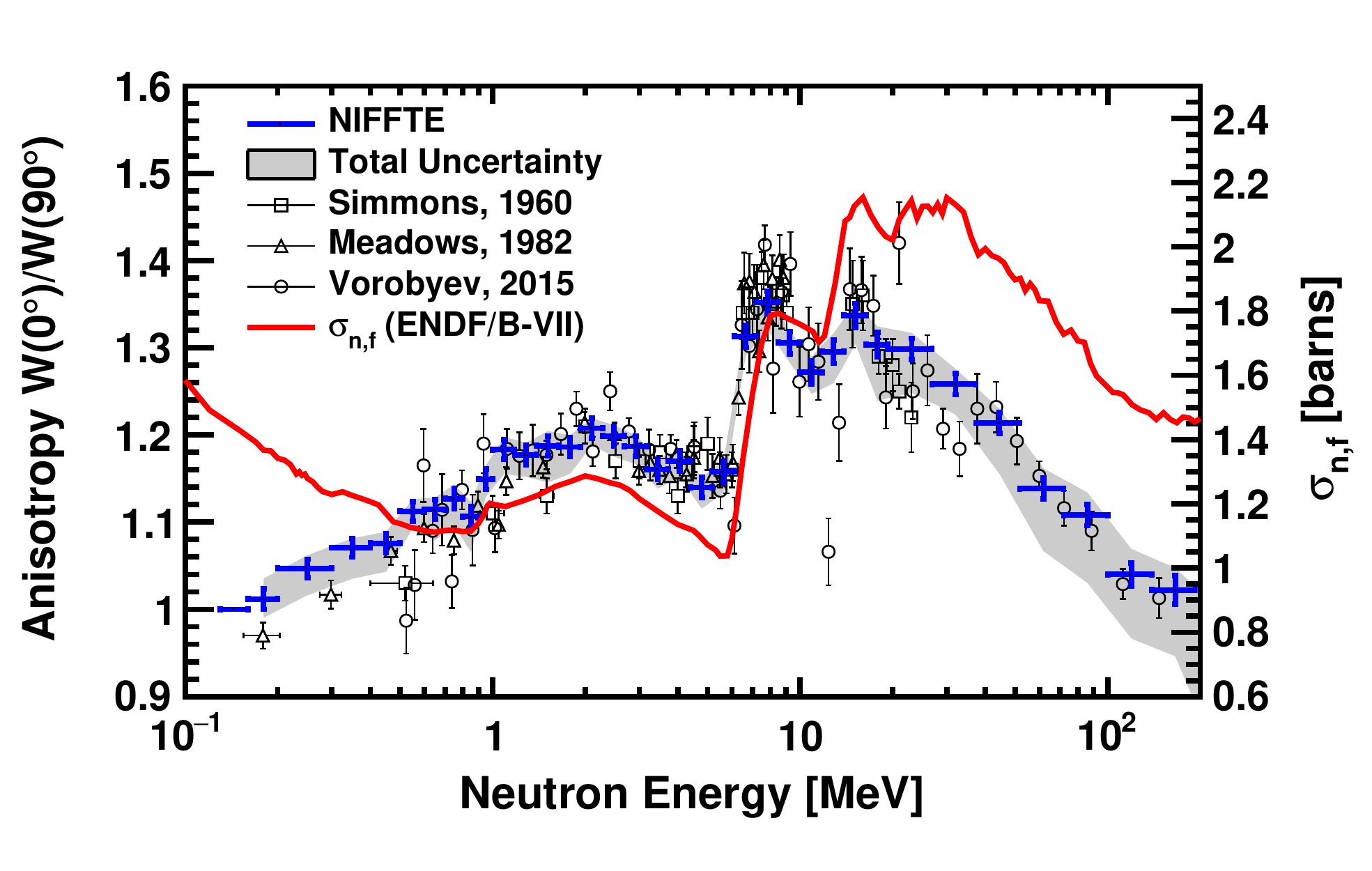}
\caption[]{(Color online) The fission fragment anisotropy as a function of neutron energy for the neutron-induced fission of $^{235}$U from 0.18 to 200~MeV compared to the most comprehensive data sets available in the literature \cite{Simmons1960,Meadows1983, Vorobyev2015} and the neutron-induced fission cross section of $^{235}$U, from ENDF/B-VII \cite{Chadwick2011a}. The result is shown with statistical error bars obtained from the uncertainty on the angular distribution fit parameters. The uncertainty band drawn in gray shade and labelled `Total Uncertainty' contains both the statistical errors plus all considered systematic uncertainties, assumed to be independent and added in quadrature.}
\label{fig:resultPlot}
\end{figure*}

Figure \ref{fig:resultPlot} shows the fission fragment anisotropy result for neutron-induced fission of $^{235}$U over neutron energies from 180~keV to 200~MeV. The tabulated result is available in the supplemental material \cite{supplement}.
The anisotropy result is shown in comparison with the three most comprehensive available data sets \cite{Simmons1960,Meadows1983,Vorobyev2015}. 

While the result agrees well with the available literature across the whole energy range, it also provides additional information on the structure of the anisotropy.
It is found that the anisotropy follows the fission cross section closely, as can be seen in Figure \ref{fig:resultPlot}. 
The result presented here especially mimics the step-like structure in the cross section, arising from multi-chance fission. The current study shows a slight decline in anisotropy followed by a sharp rise at both thresholds for second and third chance fission, 6 and 13~MeV respectively. The observation can be explained within the framework of the statistical theory of nuclear fission, as introduced in section \ref{sec:intro}. The statistical framework relates the available excitation energy to the anisotropy. With an increase in available energy, emission becomes more isotropic, while when available energy is reduced, like at the onset of multi-chance fission, emission becomes more anisotropic.

\subsection{High Energy Effects}
From 30~MeV to 200~MeV a steady decline in anisotropy is observed, as seen in Figure \ref{fig:resultPlot}.   
Agreeing with Vorobyev's measurement, our data confirms the trend of the anisotropy declining steadily from its value at 30~MeV to about 1 at incident energies beyond 200~MeV.  Within the statistical theory framework one could assume that for high neutron energies the anisotropy would eventually approach fully isotropic emission. 

For the current analysis full momentum transfer of the incoming neutron is assumed, and preequilibrium effects are therefore not separately considered. At high neutron energies, available momentum transfer data on proton-induced fission \cite{LMT} suggest additional uncertainties, limiting the current analysis to 200~MeV. Ryzhov \cite{ryzhov} discusses the effect on the anisotropy and mentions that preequilibrium processes are partly responsible for the anisotropy declining towards 1 at higher energies. It is expected that combined preequilibrium effects, multi-chance fission, and statistical distribution of quantum states at the barrier lead to a ‘washing out’ of the angular distributions - consistent with the statistical model. 
Spallation products also need to bße considered at high incident energies, but exhibit small cross sections for the  energy range in the current analysis. Spallation products are also expected to be forward peaked, their contribution is virtually eliminated by limiting the angular range of the fit and the energy threshold for the fragments. \\

\section{Conclusions}
Fission fragment angular distributions and anisotropy values for neutron-induced fission of $^{235}$U are provided in this paper over a wide incident neutron energy range, expanding the current available nuclear data into the high neutron energy range. It marks the first time that a three-dimensional tracking detector has been used to study angular distributions and anisotropy in fission fragments. The angular analysis of $^{239}$Pu data is currently in progress, and a different analysis technique without low energy normalization is under review. The angular data is expected to drive future theory developments by constraining fission process and cross section modelling. The newly developed fissionTPC instrument can trace charged particles allowing the detailed investigation of the highly complex process of nuclear fission. It has been shown that the powerful three-dimensional tracking capability makes the fissionTPC a uniquely capable tool for future nuclear fission measurements.

%
%
\begin{acknowledgments}
This work benefited from the use of the LANSCE accelerator facility and was performed under the auspices of the US Department of Energy by Los Alamos National Security, LLC under contract DE-AC52-06NA25396 and by Lawrence Livermore National Security, LLC under contract DE-AC52-07NA27344.
University collaborators acknowledge support for this work from the U.S. Department of Energy Nuclear Energy Research Initiative Project Number 08-014, the U.S. Department of Energy Idaho National Laboratory operated by Battle Energy Alliance under contract number 00043028-00116, and the DOE-NNSA Stewardship Science Academic Alliances Program, under Award Number DE-NA0002921.
\end{acknowledgments}
%


\end{document}